\documentclass[10pt,conference]{IEEEtran} 
\IEEEoverridecommandlockouts
\usepackage{cite}
\usepackage{amsmath,amssymb,amsfonts}
\usepackage{algorithmic}
\usepackage{graphicx}
\usepackage{textcomp}
\usepackage{xcolor}
\def\BibTeX{{\rm B\kern-.05em{\sc i\kern-.025em b}\kern-.08em
    T\kern-.1667em\lower.7ex\hbox{E}\kern-.125emX}}
\begin{document}

\title{Battle of the Blocs: Quantity and Quality of Software Engineering Research by Origin}

\author{\IEEEauthorblockN{Lorenz Graf-Vlachy\IEEEauthorrefmark{1}\IEEEauthorrefmark{2}}
\IEEEauthorblockA{\IEEEauthorrefmark{1}Institute of Software Engineering, University of Stuttgart, Stuttgart, Germany\\
Email: lorenz.graf-vlachy@iste.uni-stuttgart.de}
\IEEEauthorblockA{\IEEEauthorrefmark{2}TU Dortmund University, Dortmund, Germany\\
Email: lorenz.graf-vlachy@tu-dortmund.de}
}

\maketitle

\begin{abstract}
Software engineering capabilities are increasingly important to the success of economic and political blocs. This paper analyzes quantity and quality of software engineering research output originating from the US, Europe, and China over time. The results indicate that the quantity of research is increasing across the board with Europe leading the field. Depending of the scope of the analysis, either the US or China come in second. Regarding research quality, Europe appears to be lagging the other blocs, with China having caught up to and even having overtaken the US over time.
\end{abstract}


\section{Introduction}

Against the backdrop of enduring geopolitical tensions, there is a renewed interest in the technological competition between economic and political blocs, and the underlying scientific competition that drives technological development. Three of the key blocs vying for position in all fields are the United States (US), the European Union (EU), and China.

Especially the rise of China is being watched closely by the other blocs. A popular narrative is that while China has spectacularly increased the quantity of its scientific output over recent years, it remains a laggard in terms of research quality~\cite{nsb}. However, researchers have recently found that China has overtaken the US in impactful research in the aggregate~\cite{wagner}. The rise of China in science in general~\cite{nature} and these findings in particular have been widely discussed~\cite{trager}.

Given its outsized importance for economic and military superiority, the production of software is of prime interest. Software engineering is relevant not only for an increasing share of economic activity (with ``software eating the world'') but also increasingly important for the development and effective use of military equipment. Consequently, it is critical to understand which economic bloc is ahead in software engineering research.

Unfortunately, extant studies do not provide a fine-grained view on blocs' research performance in the field of software engineering. The objective of this project is thus to provide such insights into the development of software engineering research quantity and quality across blocs over time.

\section{Method}

I used data from Graf-Vlachy et al.~\cite{grafvlachy} who collected a complete set of all articles (including journal and proceedings papers but excluding, for instance, book reviews and similar items) listed in Web of Science in the category ``Computer Science -- Software Engineering''. They then manually identified all articles in any of the top 20 ``Engineering \& Computer Science -- Software Systems'' venues as listed by Google Scholar at the time of data collection.

I restricted the sample to papers published between 1999 and 2019. This ensures, on the one hand, that the universe of relevant papers is sufficiently covered, as Web of Science coverage is limited in earlier years. On the other hand, it ensures that all papers in the sample had some time to be read and cited, which is necessary to assess research quality. The number of articles in the sample of Google Scholar top 20 venues is \textit{N} = 19,368. The total number of articles in the sample is \textit{N} = 363,221.

I measured the quantity of research output for each bloc (USA, EU27, EU28 [= EU27 + UK], or China) by counting the number of articles where at least one of the authors had an address in the focal bloc.

To assess the quality of a bloc's output, I calculated a derivative of the recently proposed measure of research quality named PP-top1\%~\cite{wagner}. Specifically, I first determined for each article in the sample and for each bloc whether at least one of the authors provided an address in the focal bloc. I then established for each article whether it is in the top 5\%, top 10\%, or top 25\% of the most-cited articles in a given year. Finally, I computed the ratio of top-cited articles for each bloc and year to the number of articles one would expect from that bloc in the same year if article quality was completely random (i.e., 5\%, 10\%, or 25\% of the focal bloc's research output quantity in the focal year). The resulting metric is a measure of how many top-cited papers a bloc produced, relative to the number of top-cited papers one would expect in the absence of quality differences between articles. The metric is therefore a measure of the quality of the produced research, not of its quantity (i.e., the total number of articles) or its total impact (i.e., the total number of citations received)~\cite{wagner}.

While the measure was originally proposed primarily for the top 1\% of most-cited articles~\cite{wagner}, I calculated it only for larger buckets of most-cited articles since the number of most-cited articles is very small in the top 20 venues sample at the 1\%-level, leading to strong but uninformative fluctuations in the measure (when ignoring affected years, the results are very similar to the analyses at the 5\%-level reported below).

\section{Results}

\begin{figure}[!t]
\centering
\includegraphics[width=\columnwidth]{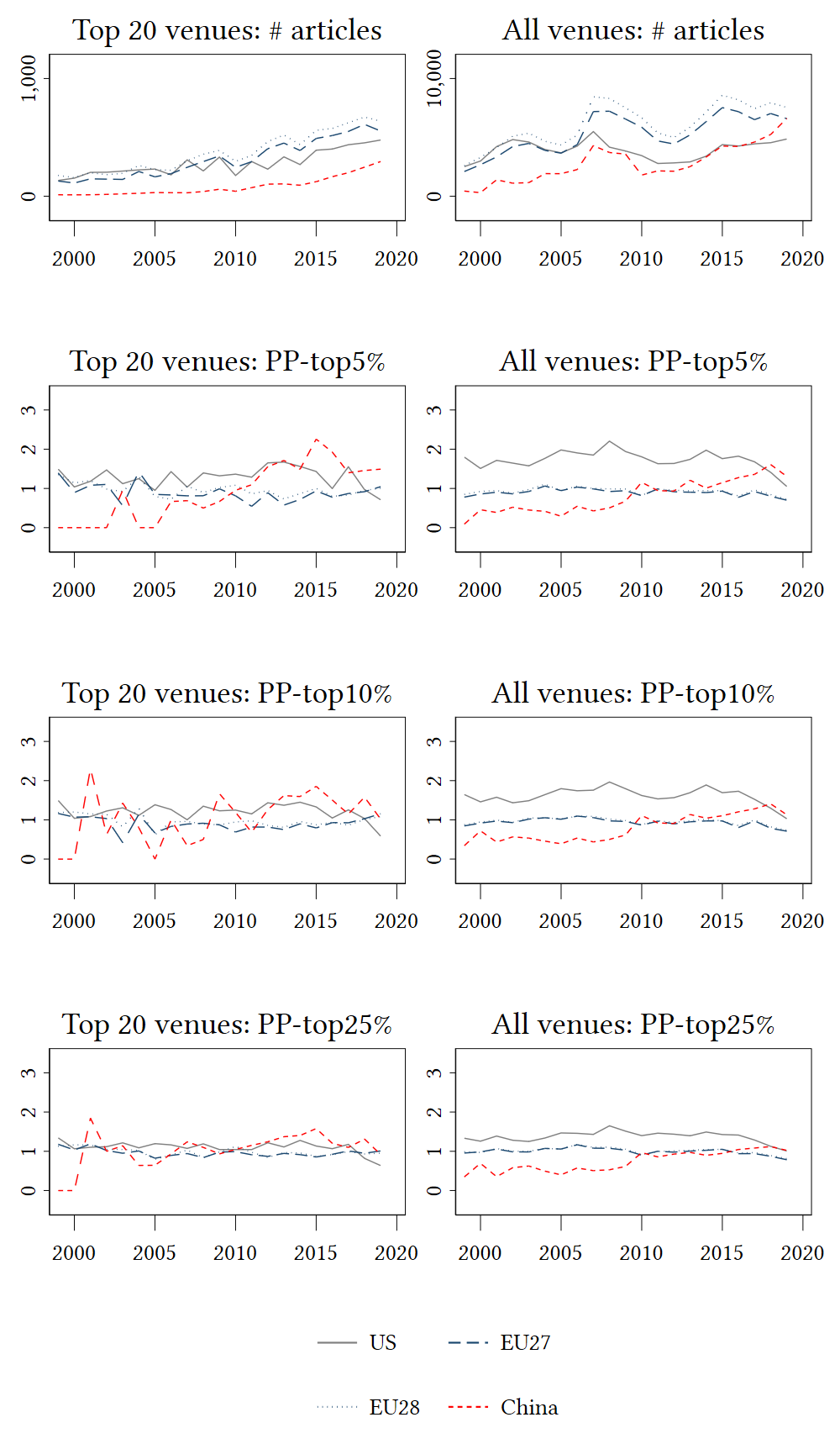}
\caption{Google Scholar top 20 software engineering venues (left) and entire Web of Science category ``Computer Science -- Software Engineering'' (right).}
\label{fig_both}
\end{figure}

Fig.~\ref{fig_both} displays the results of my analyses. The left side shows data for only the top 20 venues according to Google Scholar, whereas the right side includes data for all venues in the entire Web of Science category ``Computer Science -- Software Engineering'', which, it should be noted, also includes a substantial number of venues that are not primarily concerned with software engineering~\cite{grafvlachy}.

As is apparent, Europe is leading the pack in terms of the quantity of research output as measured by the number of publications. In both samples, Europe (as either EU27 or EU28) is ahead of both the US and China in this regard. The relationship between China and the US is more complicated because in the top 20 venues sample, China is increasing its output steadily but is also clearly and consistently lagging the US in quantity, whereas China surpassed the US in the overall sample in 2015 and essentially reached parity with the EU27 countries in 2019.

Regarding research quality, the findings are different. In the top 20 venues sample, the quality of research from both Europe and the US appears roughly flat over time, with the US showing higher research quality than Europe (at least until 2018). China, in contrast, managed to improve its software engineering research quality over the years, overtaking both Europe and the US in recent years. Perhaps unsurprisingly, this effect becomes less pronounced as I move from PP-top5\% to PP-top25\%, and thus lower the bar for what constitutes a highly-cited article.

In the overall sample, the findings regarding quality are broadly similar. Europe's performance is largely stable. The US have a similarly stable profile, with a slight decreasing trend in the most recent years. The US consistently outperform Europe in terms of software engineering research quality. China is the only bloc that shows a consistent upward trend over time, having surpassed Europe and the US. Again, this effect is more pronounced if I compute the quality measure for larger percentages of top-cited articles.

\section{Conclusion}

My analyses allow me to identify trends and draw insightful conclusions. First, it is clear that the research output in software engineering research is fluctuating over time but seems to be increasing in all blocs---particularly so in China. Depending on which set of venues one considers, China's increase is so strong that it is now on par or has even already surpassed Europe and the US. More important, however, is that China is no longer lagging the other blocs in terms of research quality. For one, China has clearly surpassed Europe in this regard. For another, depending on the venues considered, China has either caught up or even also surpassed the US regarding software engineering research quality. These trends suggest that China has reached world-class levels in software engineering, and that the dominance of the Western world in our field of research may have come to an end.

\end{document}